\documentclass[twocolumn,apj]{openjournal}

\def\msun{\hbox{${\rm M}_{\odot}$}}
\def\mstar{\hbox{${M}_{\star}$}}
\def\Zsun{\hbox{${\rm Z}_{\odot}$}}

\usepackage[breaklinks,colorlinks,citecolor=blue,urlcolor=blue]{hyperref}
\usepackage{amsmath}

\usepackage[T1]{fontenc}
\usepackage{newtxtext,newtxmath}
\usepackage{graphicx}
\usepackage{amsfonts}
\usepackage{soul}
\usepackage{hyperref}
\usepackage{enumitem}
\usepackage[hyphens]{xurl}
\usepackage{multirow}

\newcommand{\orcidauthor}[3]{\author{\href{http://orcid.org/#1}{#2$^{#3}$}}}

\renewcommand{\arraystretch}{1.2}  

\shorttitle{UMG in a $\lowercase{z} \sim 5$ Protocluster}
\shortauthors{Urbano Stawinski et al.}

\begin{document}


\title{\vspace{-0.8cm}Spectroscopic Confirmation of an Ultra-Massive Galaxy in a Protocluster at \MakeLowercase{\textit{z}}$\sim 4.9$$^{*}$\vspace{-1.5cm}}

\orcidauthor{0000-0001-8169-7249}{Stephanie M. Urbano Stawinski}{1,^\dagger}
\orcidauthor{0000-0003-1371-6019}{M. C. Cooper}{1}
\orcidauthor{0000-0001-6003-0541}{Ben Forrest}{2} 
\orcidauthor{0000-0002-9330-9108}{Adam Muzzin}{3}
\orcidauthor{0000-0001-9002-3502}{Danilo Marchesini}{4}
\orcidauthor{0000-0002-6572-7089}{Gillian Wilson}{5} 
\orcidauthor{0000-0003-0408-9850}{Percy Gomez}{6}
\orcidauthor{0000-0002-2446-8770}{Ian McConachie}{7}
\orcidauthor{0000-0002-7248-1566}{Z. Cemile Marsan}{3}
\orcidauthor{0000-0002-8053-8040}{Marianna Annunziatella}{8}
\orcidauthor{0000-0003-2144-2943}{Wenjun Chang}{7}

\affiliation{$^1$Department of Physics \& Astronomy, University of California, Irvine, 4129 Reines Hall, Irvine, CA 92697, USA}
\affiliation{$^{2}$Department of Physics and Astronomy, University of California Davis, One Shields Avenue, Davis, CA 95616 USA} 
\affiliation{${^3}$Department of Physics and Astronomy, York University, 4700, Keele Street, Toronto, ON MJ3 1P3, Canada}
\affiliation{${^4}$Department of Physics and Astronomy, Tufts University, 574 Boston Avenue, Medford, MA 02155, USA}
\affiliation{$^{5}$Department of Physics, University of California Merced, 5200 North Lake Rd., Merced, CA 95343, USA} 
\affiliation{${^6}$W.M. Keck Observatory, 65-1120 Mamalahoa Hwy., Kamuela, HI 96743, USA} 
\affiliation{${^7}$Department of Physics \& Astronomy, University of California Riverside, 900 University Ave., Riverside, CA 92521 USA} 
\affiliation{${^8}$Centro de Astrobiolog\'ia (CSIC-INTA), Ctra de Torrej\'on a Ajalvir, km 4, E-28850 Torrej\'on de Ardoz, Madrid, Spain}

\thanks{\vspace{0.1cm}$^\dagger$Corresponding author: \href{mailto:ststawinski@gmail.com}{ststawinski@gmail.com}}

\thanks{$^*$The data presented herein were obtained at the W. M. Keck Observatory, which is operated as a scientific partnership among the California Institute of Technology, the University of California and the National Aeronautics and Space Administration. The Observatory was made possible by the generous financial support of the W. M. Keck Foundation.}

\begin{abstract}\vspace{0.2cm}
    We present spectroscopic confirmation of an ultra-massive galaxy (UMG) with $\log(M_\star/M_\odot) = 10.98 \pm 0.07$ at $z_\mathrm{spec} = 4.8947$ in the Extended Groth Strip (EGS), based on deep observations of Ly$\alpha$ emission with Keck/DEIMOS. The ultra-massive galaxy (UMG-28740) is the most massive member in one of the most significant overdensities in the EGS, with four additional photometric members with $\log(M_\star/M_\odot) > 10.5$ within $R_\mathrm{proj} \sim 1$~cMpc. The Ly$\alpha$ profile is highly asymmetric ($A_f = 3.56$), suggesting the presence of neutral gas within the interstellar medium, circumgalactic medium, or via AGN-driven outflows. Spectral energy distribution (SED) fitting using a large suite of star formation histories and two sets of high-quality photometry from ground- and space-based facilities consistently estimates the stellar mass of UMG-28740 to be $\log(M_\star/M_\odot) \sim 11$ with a small standard deviation between measurements ($\sigma = 0.07$). While the best-fit SED models agree on stellar mass, we find discrepancies in the estimated star formation rate for UMG-28740, resulting in either a star-forming or quiescent system. {\it JWST}/NIRCam photometry of UMG-28740 strongly favors a quiescent scenario, demonstrating the need for high-quality mid-IR observations. Assuming the galaxy to be quiescent, UMG-28740 formed the bulk of its stars at $z > 10$ and is quenching at $z \sim 8$, resulting in a high star formation efficiency at high redshift ($\epsilon \sim 0.2$ at $z \sim 5$ and $\epsilon \gtrsim 1$ at $z \gtrsim 8$). As the most massive galaxy in its protocluster environment, UMG-28740 is a unique example of the impossibly early galaxy problem.
\end{abstract}

\section{Introduction}
The discovery of extremely massive, passive galaxies at $z \sim 3-4$ has challenged our understanding of galaxy formation and evolution at early times \citep{Schreiber_2018,Forrest_2020a,Forrest_2020b,Valentino_2020}. These quiescent ultra-massive galaxies (UMGs, $\mstar > 10^{11}~\msun$) have little to no ongoing star formation, requiring a period of intense star formation followed by rapid quenching within the first 2 Gyr of the Universe. Such extreme systems are difficult to reproduce in galaxy formation models, with current simulations struggling to form UMGs so early in cosmic time and often failing to quench these systems by $z \sim 3$ \citep{Forrest_2020a, Valentino_2023, Xie_2024}. Thus, although rare, confirmation of UMGs provide powerful constraints on theoretical models of galaxy formation and evolution \citep[e.g.][]{baugh06,croton06,nelson18}.

Intriguingly, the modeled ages and star formation histories of UMGs at $z \sim 3$ suggest that some systems assembled the majority of their mass and subsequently quenched as early as $z \sim 5$ \citep{Forrest_2020b,Schreiber_2018}. Overall, there appears to be rapid evolution of the UMG population at $4 < z < 6$, as some of these massive systems transition from highly star-forming to passive. Imaging campaigns that originally uncovered UMGs at $z \sim 3$ also corroborate these findings, yielding samples of photometric UMG candidates at $z > 4$ \citep[e.g.][]{Muzzin_2013a, Muzzin_2013b, Weaver_2023}. Meanwhile, measurements of the number density of UMGs find a significant increase in their abundance with cosmic time from $z \sim 6$ to $z \sim 3$ \citep{Stefanon_2015, Davidzon_2017, Marsan_2022, Weibel_2024}. Altogether, deep photometric surveys point to not only the existence of these extreme systems at $z \sim 5$, but also that this epoch is critical for understanding the rapid evolution of this extreme population. 

While it is presumed that ultra-massive galaxies should exist at $z \sim 5$, spectroscopic confirmation of candidate systems has been challenging. Studies searching for UMGs at $z \sim 5$ have encountered difficulty in identifying and detecting these systems via both ground- and space-based spectroscopy (\citealt{AntwiDanso_2023}; Marsan et al.~in prep). The few existing, spectroscopically-confirmed UMGs at $z > 4$ represent a biased population of systems with exceptionally high star formation rates (${\rm SFR} \gtrsim 500~\msun~{\rm yr}^{-1}$) and with relatively poorly-constrained stellar masses \citep[e.g.][but see also \citealt{Carnall_2023_N}]{Xiao_23}. 

Herein, we present the analysis surrounding the spectroscopic discovery (see \citealt{US_2024_AAS})\footnote{While finalizing this manuscript, \citet{deGraaff_2024} presented {\it JWST}/NIRSpec observations of this object, spectroscopically confirming its quiescence. These complementary and parallel results are discussed in more detail in \S\ref{subsec:Rubies}.} of an ultra-massive ($\mstar \sim 10^{11}~\msun$) galaxy at $z = 4.8947$ that resides within an extreme overdense region in the Extended Groth Strip (EGS). In \S\ref{sec:data}, we describe our target selection criterion and spectroscopic observations. We present the resulting spectral fitting and characterization of the protocluster environment in \S\ref{sec:analysis}. We discuss the competing star formation histories (SFH), compare our object to existing simulations, and investigate the mass assembly history of this object in \S\ref{sec:disc}. Finally, we summarize our results in \S\ref{sec:conc}. 
Unless otherwise specified, we adopt a $\Lambda$CDM cosmology throughout with $\Omega_{\rm M} = 0.3$, $\Omega_\Lambda = 0.70$, and $H_0 = 68$ km s$^{-1} $ Mpc$^{-1}$ \citep{PlanckCollaboration_2016_594A}.

\section{Data}\label{sec:data}

\subsection{Target Selection and Photometric Measurements}\label{sec:target}

We selected UMG candidates ($\mstar \gtrsim 10^{11}~\msun$) at $3.5 < z_{\rm phot} < 6.5$ in the Extended Groth Strip (EGS) using the multi-wavelength photometric catalog from the Cosmic Assembly Near-infrared Deep Extragalactic Legacy Survey (CANDELS; \citealt{Grogin_2011, Koekemoer_2011}). The CANDELS photometric catalog \citep{Stefanon_2017}, which spans $0.4-8\mu$m via a suite of ground- and space-based imaging,  incorporates photometric redshifts and stellar masses from seven separate photo-$z$ codes and/or spectral energy distribution (SED) template sets. For selection purposes, we used the median $z_\mathrm{phot}$ and $\mstar$ reported in the catalog. To increase the likelihood of detection of Ly$\alpha$, we prioritized objects that had an apparent CFHT $r$-band AB magnitude of $<28$ and star formation rate (SFR) of $\gtrsim 100~\msun~{\rm yr}^{-1}$. To increase our sample size, we included UMG candidates within the given redshift range that had lower SFRs and masses as filler targets. 

UMG-28740 ($\alpha = 14^\text{h}19^\text{m}39.7^\text{s}$ and $\delta = +52^\circ56^\text{m}56.5^\text{s}$), in particular, was selected as a star-forming candidate, with an estimated SFR of $20-200~\msun~{\rm yr}^{-1}$ and an apparent CFHT $r$-band magnitude of $27.9^{+0.6}_{-0.4}$. Within the \citet{Stefanon_2017} catalog, UMG-28740 has a median photometric redshift of $z_\mathrm{phot} = 4.495$ and a median stellar mass of $\log(\mstar/\msun) = 11.0$. In total, the CANDELS catalog reports photometric measurements of UMG-28740 at $ugriz$ and $JHK_s$ bands from CFHT, F606W, F814W, F125W, F140W, and F160W from {\it HST}, as well as all four {\it Spitzer}/IRAC channels. 

In addition to the photometry from CANDELS, this galaxy was recently observed with {\it{JWST}}/NIRCam \citep{Rieke_2003,Beichman_2012} as part of the Cosmic Evolution Early Release Survey (CEERS, PID 12117; \citealt{Finkelstein_2023a, Bagley_2023}). Using {\it JWST}/NIRCam (pointing 2), UMG-28740 was observed with the F115W, F150W, F200W, F277W, F356W, F410M, and F444W filters at a $5\sigma$ depth ranging from AB magnitudes of $28.35-29.2$. While the CEERS photometry was not incorporated in the initial target selection, it is utilized in this work as a supplemental set of photometry in $\S\ref{sec:SED Fit}$. 

\subsection{Spectroscopic Observations and Reduction}\label{subsec:spectra}

Using Keck/DEIMOS \citep{Faber_2003}, we observed UMG-28740 in March and May of 2023 on three different slitmasks. For all observations, we used the 600 lines mm$^{-1}$ grating blazed at 7500~\AA\ and tilted to a central wavelength of 7200~\AA, with the GG455 order-blocking filter employed and no dithering between exposures. With slit widths of $1^{\prime\prime}$, the spectral resolution (FWHM) for the 600g grating on DEIMOS is $\sim 3.5$~\AA\ \citep{Weiner_2006}, with a dispersion of $0.65$~\AA\ per pixel. Calibrations for each slitmask included three internal quartz lamp flat-field frames and an arc lamp spectrum (using Kr, Ar, Ne, and Xe lamps). We were able to observe UMG-28740 with the three different slitmasks for a combined exposure time of 6.9~hours. The average seeing for these observations ranged from $\sim 0.8$-$1^{\prime\prime}$ with variable transparency. 

\begin{figure}
    \centering
    \includegraphics[width=\linewidth]{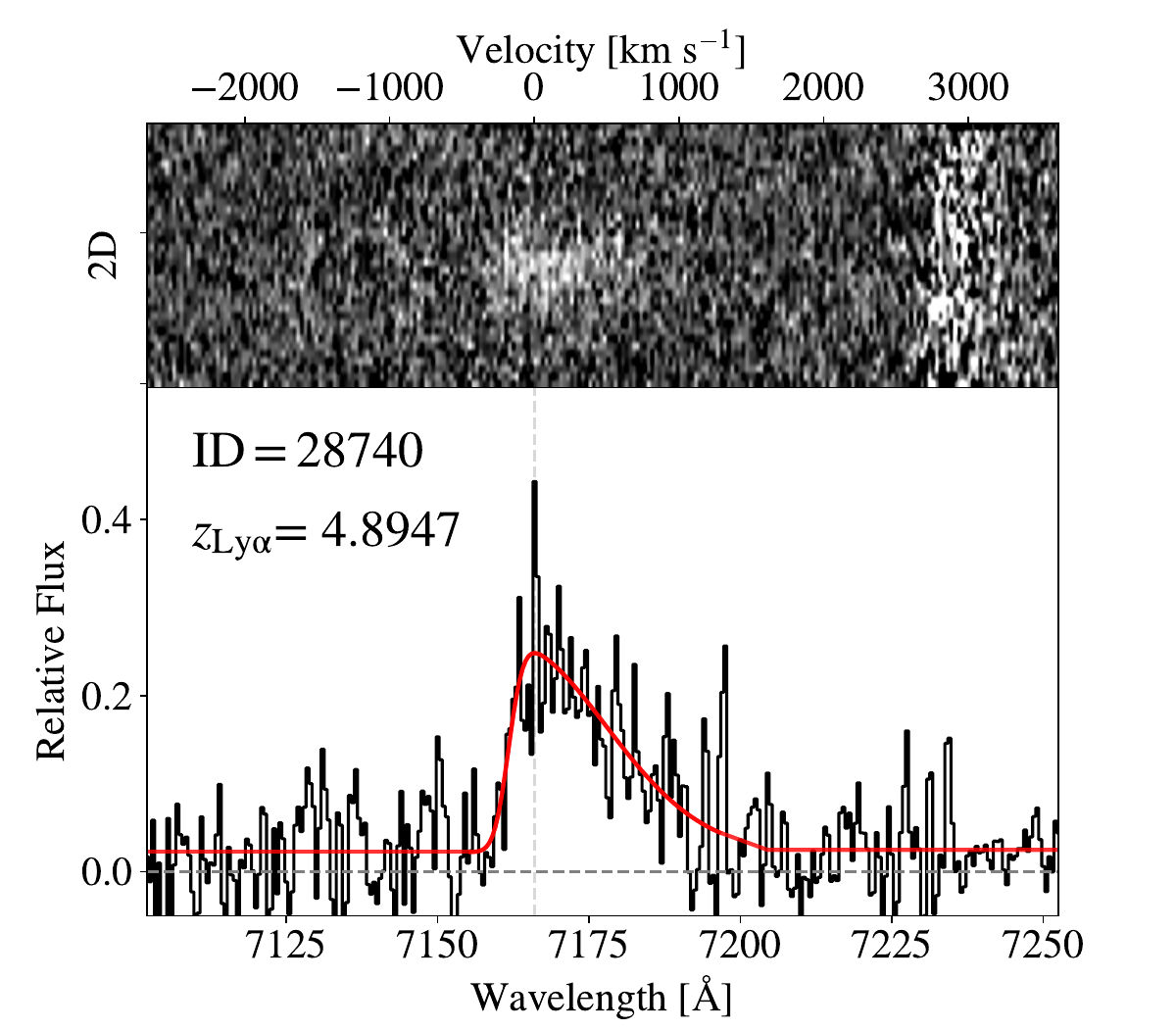}
		\caption{(\emph{top}) A roughly $150$ {\AA}-wide section of the co-added 2D spectrum for UMG-28740 from our Keck/DEIMOS observations, highlighting the Lyman-$\alpha$ emission line. (\emph{bottom}) The corresponding co-added 1D spectrum with the resulting best-fit redshift shown in red. Top $x$-axis shows the velocity offsets from $z_\mathrm{Ly\alpha}$ while the bottom $x$-axis shows the observed wavelength. 
\label{fig:spectra}}
\end{figure}

The Keck/DEIMOS observations were reduced using the \texttt{spec2d}\footnote{\url{https://sites.uci.edu/spec2d/}} DEEP2/DEEP3 DEIMOS data reduction pipeline \citep{Newman_2013,Cooper_2012_spec2d}, yielding sky-subtracted, wavelength-calibrated 1D and 2D spectra for each target. We then combined the three 1D spectra of UMG-28740 from the reduction of each slitmask, creating a final co-added 1D spectrum as shown in Figure~\ref{fig:spectra}. For visualization of the line, we also co-added the corresponding 2D spectra, as shown in the top panel of Figure~\ref{fig:spectra}. 

To measure a best-fit redshift, we utilized a custom template fitter (as used in \citealt{US_2024}) that incorporates both an emission-line galaxy template as well as an asymmetric Gaussian profile to probe for a single Ly$\alpha$ line. For UMG-28740, we found a single emission line nearby the expected wavelength for Ly$\alpha$ given the photometric redshift. Fitting this profile, we find the resulting heliocentric-corrected spectroscopic redshift for UMG-28740 is $z_\mathrm{spec} = 4.8947$ (best-fit model is shown by the red line in Fig.~\ref{fig:spectra}). 

While this redshift is determined by the presence of a single emission line, there are multiple arguments consolidating this feature to be Ly$\alpha$. First, the asymmetric profile of the emission line follows the characteristic shape of Ly$\alpha$ emitters at $z > 3$ \citep[e.g.][]{Verhamme_2008, Martin_2015}. Second, the measured redshift is consistent with two sets of photometry from CANDELS and CEERS, including high-quality photometric measurements from {\it HST}/WFC3 and {\it JWST}/NIRCam (see Fig. \ref{fig:target_stamp}). Lastly, the location of this galaxy coincides with a large overdense region at $z \sim 5$ (first highlighted by \citealt{Naidu_2022} and later discussed in \S\ref{subsec:overdensity}), which further solidifies $z_\mathrm{spec} = 4.8947$ as the correct redshift. 

The asymmetric profile of Ly$\alpha$ suggests the presence of neutral gas around this object. We quantify the asymmetry of the Ly$\alpha$ emission by taking the ratio of the red and blue sides of the line profile following $A_f = (\int^{\infty}_{\lambda_\mathrm{peak}} f_\lambda \ d\lambda) / (\int^{\lambda_\mathrm{peak}}_{\lambda_\mathrm{trough}} f_\lambda \ d\lambda) $ where $f_\lambda$ is the flux, $\lambda_\mathrm{peak}$ is the wavelength center of the profile, and $\lambda_\mathrm{trough}$ is the blue edge \citep{Kakiichi_2021,Rhoads_2003}. We find the asymmetry parameter to be high, $A_f = 3.56$, potentially indicating sufficient neutral gas surrounding the galaxy within the interstellar medium, the circumgalactic medium, or possibly as part of AGN-driven outflows \citep{Verhamme_2008, Laursen_2009}.

\section{Analysis}\label{sec:analysis}

\subsection{SED Fitting}\label{sec:SED Fit}

\begin{figure*}
    \centering
    \includegraphics[width=0.24\linewidth]{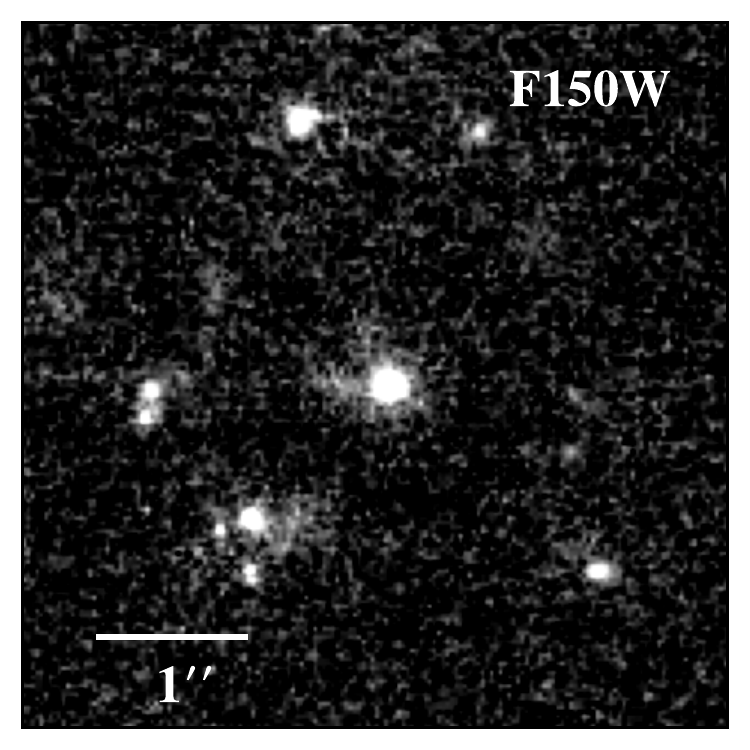}
        \includegraphics[width=0.24\linewidth]{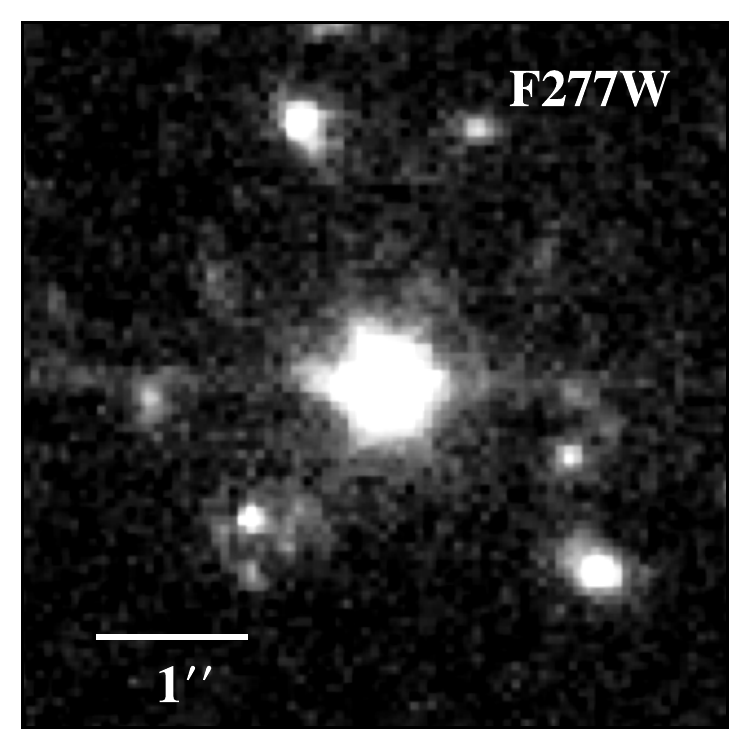}
        \includegraphics[width=0.24\linewidth]{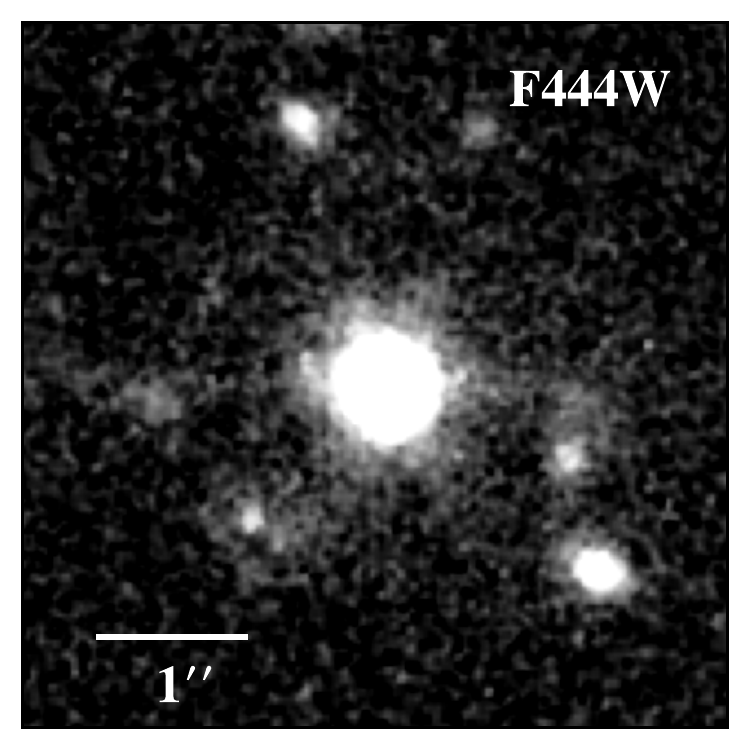}
        \includegraphics[width=0.24\linewidth]{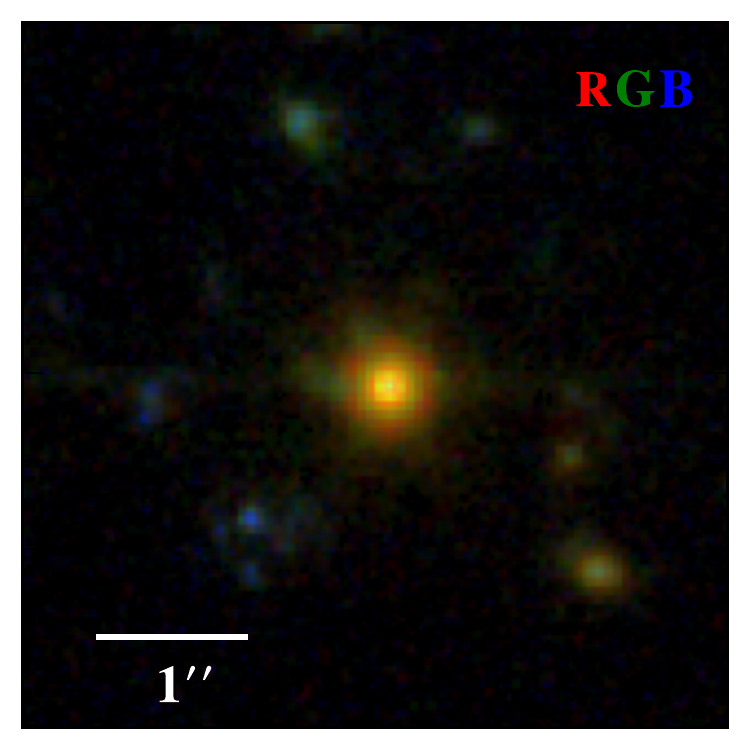}
    \includegraphics[width=\linewidth]{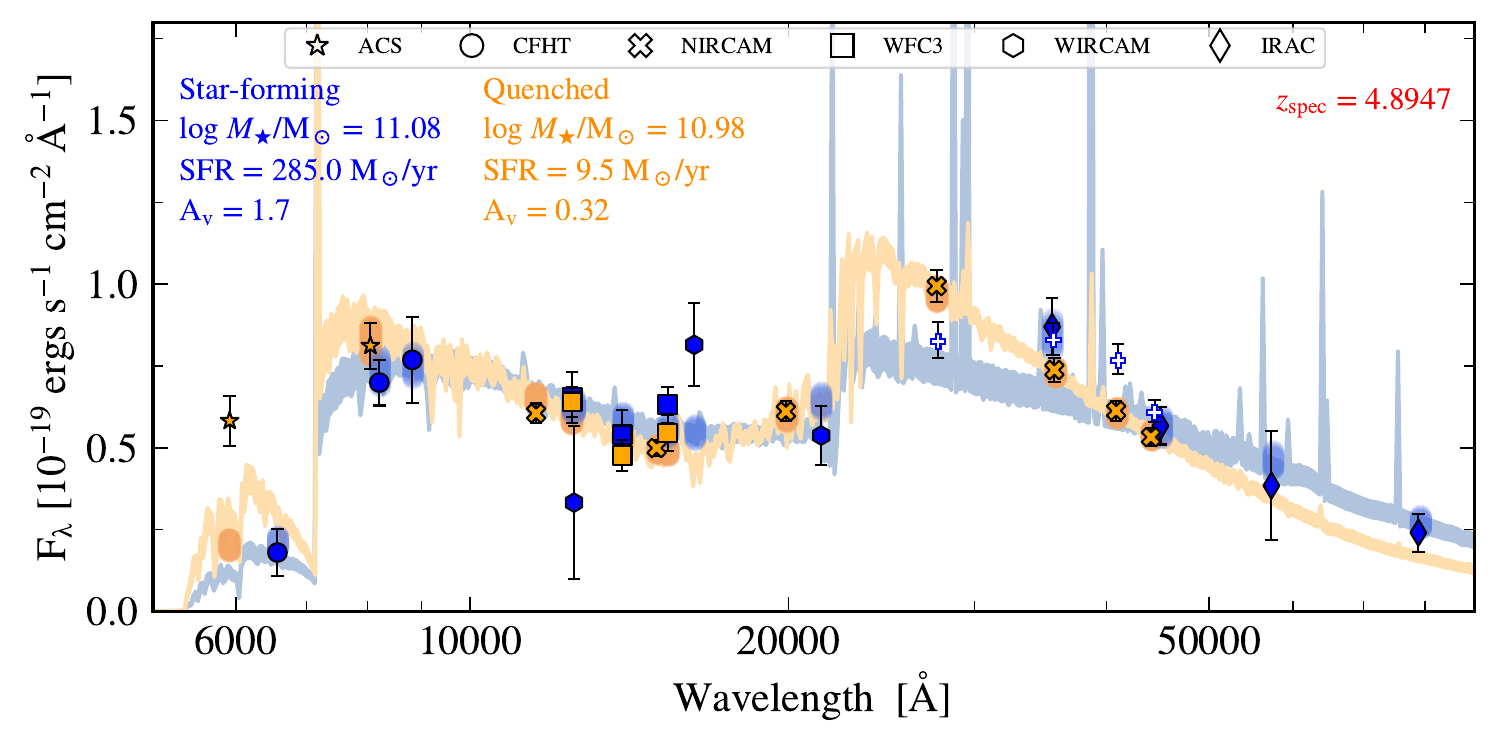}
		\caption{(\emph{top}) From left to right: image stamps of UMG-28740 in the {\it{JWST}}/NIRCam 1.50$\mu$m, 2.77$\mu$m, and 4.44$\mu$m filters, followed by an RGB image (F150M in blue, F277M in green, and F444W in red). (\emph{bottom}) Best-fit SEDs (star-forming in blue and quenched in orange) from \texttt{BAGPIPES} forced at the measured $z_\mathrm{spec}$. The two sets of observed photometry are shown as dark blue \citep{Stefanon_2017} and orange \citep{Bagley_2023} markers. The width of each SED indicates the $\pm1\sigma$ confidence intervals resulting from \texttt{BAGPIPES}. The open blue crosses show the predicted {\it JWST}/NIRCam flux at F277W, F356W, F410M, and F444W for the blue star-forming SED. Here, the predicted {\it JWST}/NIRCam flux for the star-forming scenario is under-estimated at $2.77\mu{\rm m}$ and over-estimated at $3.56-4.44\mu{\rm m}$, suggesting that UMG-28740 is likely quenched.  
  \label{fig:target_stamp}}
\end{figure*}
 
With our measured spectroscopic redshift ($z_\mathrm{spec} =4.8947$), we re-fit the SED of UMG-28740. Here, we employ both the photometry from CANDELS \citep{Stefanon_2017} and from CEERS \citep{Bagley_2023}. From the CANDELS photometric catalog, we exclude the {\it HST}/ACS photometry at $F606W$ and $F814W$ due to uncertainties related to sky subtraction, such that the SED fits are made to the photometry at $ugrizJHK_{s}$ from CFHT, F125W, F140W, and F160W from {\it HST}/WFC3-IR, and all four {\it Spitzer}/IRAC channels. Meanwhile, the CEERS catalog includes photometric measurements at F606W and F814W from {\it HST}/ACS, F125W, F140W, and F160W from {\it HST}/WFC3-IR, and finally F115W, F150W, F200W, F277W, F356W, F410M, and F444W from {\it JWST}/NIRCam. 

The majority of our SED analysis was completed using the SED-fitting code Bayesian Analysis of Galaxies for Physical Inference and Parameter EStimation or \texttt{BAGPIPES} \citep{Carnall_2018}.\footnote{\href{https://github.com/ACCarnall/bagpipes}{https://github.com/ACCarnall/bagpipes}} \texttt{BAGPIPES} assumes Bruzal and Charlot stellar populations \citep{BC_2003}, a Kroupa IMF \citep{Kroupa_2002}, and a Calzetti dust law \citep{Calzetti_2000}. Adopting a double-power law star formation history, we measure best-fit model SEDs for each set of photometry (from CANDELS and CEERS). We fit these datasets separately to avoid any systematic biases associated with the varying photometric methods, such as different aperture sizes and adopted PSF-matching methodology. 
The SEDs are fit to 10 free parameters, allowing for variation in metallicity, age, the slope of the Calzetti dust law, and the ionization parameter. 
The full set of free parameters used and their allowed distributions are outlined in Table~\ref{table:BAGPIPESParameters}.

\begin{table}
    \centering
    \caption{List of free parameters used in the SED fits with \texttt{BAGPIPES}}
    \begin{tabular}{cccc}
    \hline
    Parameter & Unit & Range & Prior \\ \hline \hline
    \multicolumn{4}{c}{General\textsuperscript{a}} \\
    \hline 
     Stellar Mass Formed & $\mstar/\msun$ & (1, 15) & Logarithmic \\ 
     Peak Star Formation & $\tau/ $Gyr & (0.3, $t_\mathrm{obs}$) & Logarithmic \\ 
     Age & $t_\mathrm{age}/$Gyr & (0.1, 1) & Logarithmic \\ 
     Metallicity & $Z/\Zsun$ & (0.01, 2.5) & Logarithmic \\ 
     
     $V$-band attenuation & $A_V$ & (0, 8) & Uniform \\ 
     Deviation from & \multirow{2}*{$\delta$} & \multirow{2}*{(-0.3, 0.3)} & \multirow{2}*{Gaussian\textsuperscript{b}} \\ 
          Calzetti slope &  & &  \\ 
          
     2175\AA \ bump strength & $B$ & (0, 5) & Uniform \\ 
     Ionization & $U$ & (-4, -2) & Logarithmic \\ \hline
    \multicolumn{4}{c}{Double-power law\textsuperscript{c}} \\
    \hline 
     double-power law & \multirow{2}*{$\alpha $ } & \multirow{2}*{(0.01, 1000)} & \multirow{2}*{Logarithmic} \\
      falling slope & &  &  \\ 
     
         double-power law & \multirow{2}*{$\beta $} & \multirow{2}*{(0.01, 1000)} & \multirow{2}*{Logarithmic} \\ 
      rising slope &  &  &  \\ \hline

    \multicolumn{4}{c}{Burst\textsuperscript{d}} \\ \hline 
     Stellar Mass Formed & $\mstar/\msun$ & (0, 15) & Logarithmic \\ 
     Age & $t_\mathrm{age}/$Gyr & (0.1, 1) & Logarithmic \\ 
     Metallicity & $Z/\Zsun$ & (0.01, 2.5) & Logarithmic \\ \hline  
     
    \multicolumn{4}{c}{AGN\textsuperscript{d}} \\ \hline 
     Continuum Flux at & $f_{5100}$ & \multirow{2}*{(0, $10^{-19}$)} & \multirow{2}*{Uniform} \\ 
          5100\AA & erg/s/cm$^{2}$/\AA &  & \\ 
    \multirow{2}*{H$\alpha$ Flux} & $f_\mathrm{H\alpha}$  & \multirow{2}*{(0, $10^{-16.6}$)} & \multirow{2}*{Uniform} \\
         &  erg/s/cm$^{2}$  &  & \\
    H$\alpha$ Velocity  & \multirow{2}*{$\sigma_\mathrm{H\alpha}$/km/s } & \multirow{2}*{(1000, 5000)} & \multirow{2}*{Logarithmic} \\
 Dispersion &  &  &  \\
Power law slope: & & & \\
      at $\lambda < 5000$\AA & $\alpha_{\lambda < 5000}$ & (-2, 2) & Gaussian\textsuperscript{e} \\ 
      at $\lambda > 5000$\AA & $\alpha_{\lambda > 5000}$ & (-2, 2) & Gaussian\textsuperscript{f} \\ \hline 
     
     \multicolumn{4}{l}{\textsuperscript{a}\footnotesize{Parameters used in all SED fits.}}\\
     \multicolumn{4}{l}{\textsuperscript{b}\footnotesize{The Gaussian distribution for the Calzetti slope has a mean of $\mu = 0$  }}\\
     \multicolumn{4}{l}{and a standard deviation of $\sigma =0.1$}\\
     \multicolumn{4}{l}{\textsuperscript{c}\footnotesize{Parameters specifically used in the double-power law fits. }}\\
     
     \multicolumn{4}{l}{\textsuperscript{d}\footnotesize{Parameters used in added AGN and burst components.}}\\
     \multicolumn{4}{l}{\textsuperscript{e}\footnotesize{The Gaussian distribution for the power law slope at at $\lambda < 5000$\AA  }}\\
     \multicolumn{4}{l}{has a mean of $\mu = -1.5$ and a standard deviation of $\sigma =0.5$}\\
     \multicolumn{4}{l}{\textsuperscript{f}\footnotesize{The Gaussian distribution for the power law slope at at $\lambda > 5000$\AA  }}\\
     \multicolumn{4}{l}{has a mean of $\mu = 0.5$ and a standard deviation of $\sigma =0.5$}\\
    \end{tabular}
    \label{table:BAGPIPESParameters}
\end{table}

\begin{table}
    \centering
    \caption{Stellar Mass and SFR from Alternative SED Fits}
    \begin{tabular}{cccc}
    \hline
    SFH & Code & Mass & SFR \\
     &  & $\log(\mstar/\msun)$ & $\msun$/yr \\\hline \hline
    \multicolumn{4}{c}{CANDELS \citep{Stefanon_2017}} \\
    \hline 
    Double-power law\textsuperscript{a} \vspace{2mm} & \texttt{BAGPIPES} & 11.08 $^{ 0.08 }_{ 0.1 }$ & 285.0 $^{ 125.5}_{ 79.5 }$  \\
    Double-power law+AGN \vspace{2mm} & \texttt{BAGPIPES} & 10.97 $^{ 0.04 }_{ 0.04 }$ &  8.6 $^{ 3.9 }_{ 2.4 }$\\
    Double-power law+burst \vspace{2mm} & \texttt{BAGPIPES} & 10.98 $^{ 0.08 }_{ 0.09 }$ &  419.6 $^{ 156.7 }_{ 162.2 }$\\
    Double-power law at $Z_\odot$ \vspace{2mm} & \texttt{BAGPIPES} & 10.91 $^{ 0.1 }_{ 0.07 }$  &  299.9 $^{ 99.5 }_{ 60.1 }$ \\
    Delayed-$\tau$ \vspace{2mm} & \texttt{BAGPIPES} & 10.92 $^{ 0.11 }_{ 0.13 }$ & 312.1 $^{ 159.1 }_{ 106.2 }$  \\
    Double Exponential & \texttt{FAST++} & 10.95$^{ 0.00 }_{ 0.11 }$ & 24.5$^{ 16.6 }_{ 3.3 }$
    \\ \hline 
    \multicolumn{4}{c}{CEERS \citep{Bagley_2023}} \\
    \hline 
    Double-power law\textsuperscript{a}  \vspace{2mm}& \texttt{BAGPIPES} & 10.98 $^{ 0.04 }_{ 0.03 }$ & 9.5 $^{ 4.8 }_{ 2.7 }$  \\
    Double-power law+AGN \vspace{2mm} & \texttt{BAGPIPES} & 10.96 $^{ 0.04 }_{ 0.04 }$ &  0.0 $^{ 0.0 }_{ 0.0 }$ \\
    Double-power law+burst \vspace{2mm} & \texttt{BAGPIPES} & 11.17 $^{ 0.03 }_{ 0.04 }$ &  18.1 $^{ 6.2 }_{ 4.7 }$ \\
    Double-power law at $Z_\odot$ \vspace{2mm} & \texttt{BAGPIPES} & 10.99 $^{ 0.02 }_{ 0.02 }$ &  7.4 $^{ 1.0 }_{ 0.7 }$  \\
    Delayed-$\tau$ & \texttt{BAGPIPES} \vspace{2mm} & 11.05 $^{ 0.02 }_{ 0.03 }$  & 32.2 $^{ 10.8 }_{ 5.7 }$ \\ \hline
    \multicolumn{4}{l}{\textsuperscript{a}\footnotesize{SED fit is shown in Fig. \ref{fig:target_stamp}.}}
    \end{tabular}
    \label{table:SEDFits}
\end{table}

As shown in Figure~\ref{fig:target_stamp}, the resulting SED fits give a stellar mass of $\log(\mstar/\msun) = 10.98^{+0.04}_{-0.03}$ using the CEERS photometry and $\log(\mstar/\msun) = 11.08^{+0.08}_{-0.1}$ using the CANDELS photometry.
While the stellar mass is tightly constrained between the two sets of photometry, we find that the SFR is dependant on the photometric catalog utilized. The CANDELS photometry yields a dusty, highly star-forming galaxy (${\rm SFR}=285.0^{+125.5}_{-79.5}~\msun~{\rm yr}^{-1}$) and the fit to the {\it JWST}/NIRCam photometry produces a very low star formation rate (${\rm SFR}=9.5^{+4.8}_{-2.7}~\msun~{\rm yr}^{-1}$) and low dust attenuation. 
The latter, \textit{nearly} quenched SED is shown in orange in Figure~\ref{fig:target_stamp}, and is consistent with both sets of photometry within $\pm 1 \sigma$ error. On the other hand, the star-forming SED (shown in blue in Fig.~\ref{fig:target_stamp}) has insufficient emission in the rest-frame optical ($\sim4000$-$6000$\AA), so as to under-produce the flux measured in the {\it JWST}/NIRCam F277W band. 
In addition, the star-forming SED overpredicts the observed flux in the {\it JWST}/NIRCam F356W, F410M, and F444W bands due to contribution from strong emission lines such as H$\alpha$ at $\sim 3.9\mu{\rm m}$. 
Based on the photometry alone, this favors the quenched scenario for UMG-28740 (see \S\ref{sec:SFH} for a more detailed discussion of these discrepancies).

To test the robustness of our measurements for mass and SFR, we ran \texttt{BAGPIPES} again using a variety of different models; varying the included free parameters, changing the assumed star formation history, restricting the assumed metallicity to solar, and adding different components such as bursts and the broad-line region of an active galactic nucleus (AGN; modeled by Gaussian fits to the H$\alpha$ and H$\beta$ broad lines, see \citealt{Carnall_2023_N}). In summary, we highlight the results from five different models: a double-power law SFH, a double-power law SFH including an AGN, a burst, or fixing the metallicity to solar metallicity, and a delayed-$\tau$ model (see Table \ref{table:SEDFits}). Adding an AGN or burst component is of interest to the evolution of massive galaxies in the early Universe, as the spectra of a subset of UMGs have significant AGN contributions (\citealt{Marsan_2017}; \citealt{Forrest_2020b}, Forrest et al. in prep.) and current theoretical models speculate that the rapid formation of such systems require extreme bursts of star formation \citep{Sun_2023}. 

We also compared the results from \texttt{BAGPIPES} to fits using the C++-adapted SED fitting code \texttt{FAST++}\footnote{\href{https://github.com/cschreib/fastpp}{https://github.com/cschreib/fastpp}} \citep{Kriek_2009}. With \texttt{FAST++}, we adopt the same parameters used in the literature to model UMGs at $z \sim 3.5$ \citep{Forrest_2020a,Forrest_2020b}, assuming a Bruzal Charlot stellar population \citep{BC_2003}, the Calzetti dust law \citep{Calzetti_2000}, and a Chabrier IMF \citep{Chabrier_2003}. The Chabrier IMF is related to the Kroupa IMF used in \texttt{BAGPIPES} by a factor of $\sim 1$ \citep{Wright_2017}, allowing for direct comparison between results of the two SED codes. We allow \texttt{FAST++} to vary the extinction from $0 < A_V < 4$ with a $\Delta A_V = 0.1$, the age from $7 < \log(\mathrm{Age}) <$ age of the Universe at $z_\mathrm{spec}$ with $\Delta \log(\mathrm{Age}) = 0.05$, and a metallicity of $Z/\Zsun = 0.02$ as in \citet{Forrest_2020a,Forrest_2020b}. For the assumed SFH, we adopt a double-exponential SFH as used in \citet{Forrest_2020a,Forrest_2020b}. 

Remarkably, regardless of the photometric dataset used, SFH assumed, underlying SED code, or range of free parameters, we find \textit{all} best-fit SEDs result in a massive galaxy with a stellar mass of $\log(\mstar/\msun) \sim 11$. Across the various fits, UMG-28740 has a median mass of $\log (\mstar/\msun) = 10.98$ with a $1\sigma$ standard deviation of $0.07$ (see Table~\ref{table:SEDFits} for measurements from each SED fit). Hereinafter, we conservatively use this stellar mass and error as our best estimate for UMG-28740. This SED fitting analysis also highlights a discrepancy between SFR and dust extinction -- resulting in the two possible scenarios, star-forming or quenched (see \S\ref{sec:SFH} for a more detailed discussion).

\subsection{An Overdense Region around UMG 28740}\label{subsec:overdensity}

\begin{figure}
    \centering
    \includegraphics[width=\linewidth]{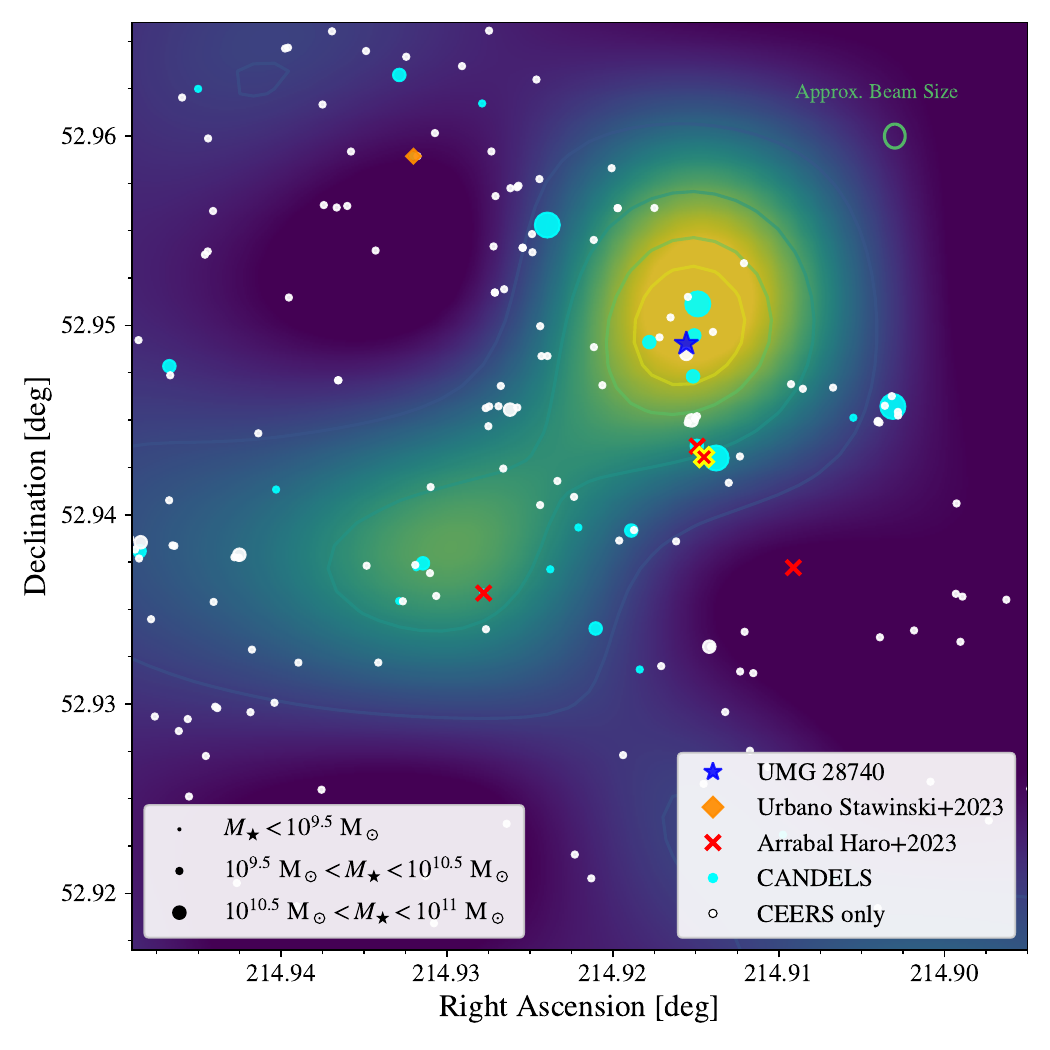}
		\caption{Overdensity hosting UMG-28740 (blue star) at $z \sim 5$. The underlying 2D histogram and contours show the galaxy surface density for objects in the CANDELS catalog at $4.5 < z_\mathrm{phot} < 5.5$ and $\log (\mstar/\msun) > 9$. The beam size (with a diameter of $4.5^{\prime\prime}$) used to convolve the density map is shown in the upper right corner. Cyan circles denote photometric sources found in the CANDELS \citep{Stefanon_2017} and CEERS \citep{Bagley_2023} catalogs, where white circles are low-mass galaxies only found in the deeper CEERS catalog. The size of the circle indicates the estimated stellar mass of the object. Red crosses and the orange diamond are spectroscopically-confirmed sources from \citet{Arrabal_2023Natur} and \citet{US_2024}, respectively. The highlighted red cross indicates the location of the lower-mass member from \citet{Naidu_2022}. 
\label{fig:density_map}}
\end{figure}

UMG-28740 is embedded within a significant overdensity at $z \sim 5$, as traced by the surface density of galaxies at $4.5 < z_{\rm phot} < 5.5$ in the CANDELS photometric catalog. As shown in Figure~\ref{fig:density_map}, UMG-28740 resides near the center of the overdense region and is the most massive member. However, we caution that the observed overdensity is located near the edge of the CANDELS survey area within the EGS, with UMG-28740 only $\sim 16^{\prime\prime}$ from the edge of the {\it HST}/WFC3 imaging footprint, such that the full extent of the structure may be underestimated. \citet{Naidu_2022} first identified this particular overdense region, finding that a $\sim 0.8~{\rm arcmin}^{2}$ area centered on a lower-mass photometric neighbor of UMG-28740 contained $4 \times$ the average surface density of sources at $4.5 < z_{\rm phot} < 5.5$. However, this lower-mass neighbor ($\mstar \sim 10^{9.6}~\msun$, spectroscopically confirmed by \citealt{ArrabalHaro_2023}), is located $\gtrsim 20^{\prime\prime}$ to the south ($\gtrsim 0.8~{\rm cMpc}$ at $z = 5$) of UMG-28740 and outside of the apparent center of the overdense region as traced by the CANDELS photometric catalog (see the highlighted red cross in Fig.~\ref{fig:density_map}). 

To investigate the extent of the overdense region around UMG-28740, we search for other overdensities across the entire CANDELS footprint within the EGS field by comparing the galaxy surface densities in redshift slices of $\Delta z = 1$ to the average surface density at each corresponding epoch. We use the CANDELS catalog to identify sources at $2.5 < z_\mathrm{phot} < 6.5$, as this catalog covers a larger footprint within the EGS (versus CEERS). We limit this analysis to sources with $\log (\mstar/\msun) > 9$, so as to define a tracer population that is unbiased (complete) over the full redshift range probed. Across the entire field, we compute the galaxy surface density as a function of position on the sky within an aperture of $45^{\prime\prime} \times 45^{\prime\prime}$ and a depth of $\Delta z = 1$. Stepping through redshift from $z=3$ to $z=6$ with a step size of $\Delta z = 0.1$, we identify regions that are overdense compared to the average surface density at the corresponding redshift. We find that the region hosting UMG-28740 is one of the most extreme overdense regions within the EGS at $3 < z < 6$, with a surface density that is $\sim 10 \times$ higher than the average surface density of the field at $z = 4.9$. 

At $z\sim5$, the effective radii of protoclusters are estimated to extend up to $R_e = 5-10$ cMpc \citep{Chiang_2013}. UMG-28740 is positioned nearby the apparent center of the overdense region, such that we can estimate $R_\mathrm{proj}$ from UMG-28740 to trace the effective radius of the protocluster. However, we caution this region is also near the edge of the EGS field, limiting observed galaxies past $R_\mathrm{proj} \sim 0.6$ cMpc to the north-west and potentially skewing the true center of the overdensity. Nevertheless, within $R_\mathrm{proj} = 3.44$~cMpc from UMG-28740, there are 26 photometric protocluster members at $4.5 < z < 5.5$ and with $\mstar > 10^{9}~\msun$ in the CANDELS photometric catalog along with 160 members identified in the CEERS catalog, pushing down to lower stellar masses ($\mstar > 10^{8}~\msun$). Additionally, out of all the photometric members there are $4$ massive neighbors with $10.5 < \log(\mstar/\msun) < 11$  within just 29.5$^{\prime\prime}$, or $\sim 1.13$ cMpc, of UMG-28740. The closest of these massive members is only separated from UMG-28740 by $0.29$~cMpc. This makes the protocluster surrounding UMG-28740 the most compact overdensity of massive $\log(\mstar/\msun)>10.5$ galaxies within the entire EGS at $z \sim 5$ (according to the CANDELS catalog), with only two other compact regions each containing two similarly massive galaxies at a similar spatial scale ($R_\mathrm{proj} < 3$ cMpc). Moreover, there are \textit{no} other instances of $>3$ massive galaxies within $R_\mathrm{proj} < 3$ cMpc at this redshift in the CANDELS catalog. 

\begin{figure*}
    \centering
    \includegraphics[width=\linewidth]{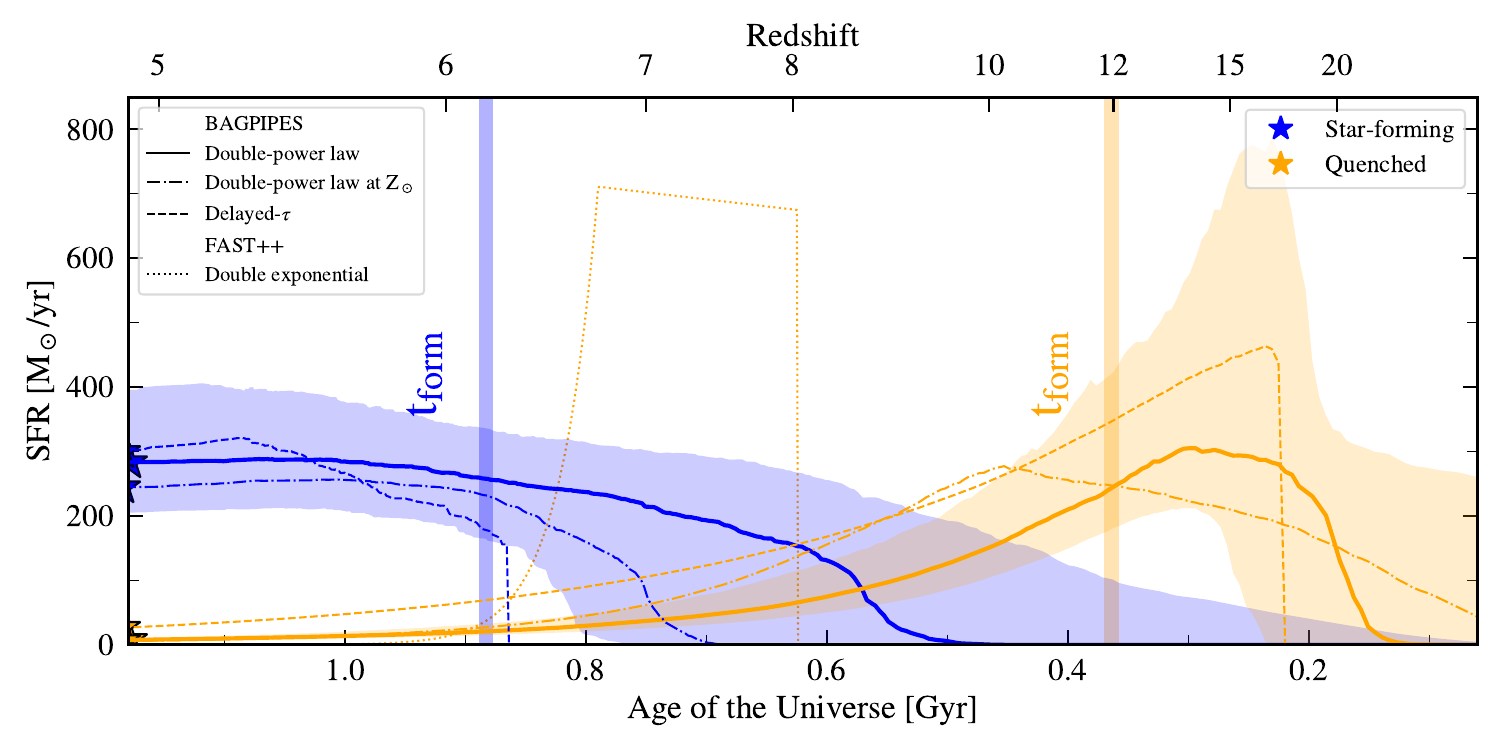}
		\caption{Star formation histories for the two competing scenarios, star-forming (blue) and quenched (orange), based on the best-fit SEDs from \texttt{BAGPIPES} and \texttt{FAST++} (see Table~\ref{table:SEDFits}). The solid lines, as well as the shaded region, show the best-fit and $\pm 1 \sigma$ double-power law SFH from \texttt{BAGPIPES} (shown in Fig. \ref{fig:target_stamp}). The formation time ($t_{\rm form}$) corresponds to the epoch at which 50\% of the stellar mass is formed for the double-power law models. The higher-quality {\it JWST}/NIRCam photometry favors the quenched scenario and suggests that UMG-28740 formed much of its stellar mass as early as $z \gtrsim 10$. 
\label{fig:competing_SFH}}
\end{figure*}

This protocluster also hosts 5 spectroscopically-confirmed galaxies from Keck/DEIMOS \citep{US_2024} and {\it JWST}/NIRSpec \citep{Arrabal_2023Natur} spectroscopy within $R_\mathrm{proj} = 3.44$ cMpc (see Fig.~\ref{fig:density_map}). The measured mass of these spectroscopic members range from $\log(\mstar/\msun) = 9.18-9.95$ and span 0.75~cMpc to 2.10~cMpc from UMG-28740. In summary, all members within $R_\mathrm{proj} = 4$~cMpc, both spectroscopically confirmed and all photometric candidates, are estimated to be less massive than UMG-28740.

\section{Discussion}\label{sec:disc}

\subsection{Investigation into Competing Star Formation Histories}\label{sec:SFH}

As mentioned in \S\ref{sec:SED Fit}, the two sets of photometry yield two different SFH scenarios, one with a highly star-forming and dusty SED versus one with low star formation and little dust extinction. The high-quality {\it JWST}/NIRCam photometry favors a nearly quenched, SED. In contrast, the photometry from CANDELS favors a star-forming SED, with the exception of a quenched solution when assuming a double-exponential SFH with \texttt{FAST++} (as described in \S\ref{sec:SED Fit}). We show a variety of these SFHs, highlighting the best fit SEDs from the final \texttt{BAGPIPES} run, in Figure \ref{fig:competing_SFH}.  

Looking at the SED fits alone, the $\pm 1 \sigma$ spread of the orange, quenched SED is consistent with both sets of photometry within the errors. However, there is a clear offset between the {\it JWST}/NIRCam photometry and the blue, star-forming SED, with the largest discrepancy at F277W and F410M. To test the consistency between the NIRCam photometry and the star-forming SED, we predict the flux that we would expect from a mock F277W, F356W, F410M, and F444W observation of the star-forming SED fit, using the `predict' function within the \texttt{BAGPIPES} code (see the open blue crosses in Fig. \ref{fig:target_stamp}). For F277W, the expected flux would be elevated from the continuum due to the emission lines nearby $2.77\mu{\rm m}$; however, we find this predicted flux is insufficient to reproduce the observed F277W flux even within $\pm 1 \sigma$. For F356W and F410M, the predicted flux is elevated to a greater extent, due to contributions from strong emission lines such as H$\alpha$ at $\sim 3.9\mu{\rm m}$. However, this added emission over-predicts the flux measured by the NIRCam photometry (also seen, to a lesser extent, in the F444W predicted flux). Taken together, this suggests that the quenched scenario is more consistent with the observed photometry from both catalogs, and thus more likely to be the true SFH of UMG-28740.  
Moreover, running \texttt{BAGPIPES} with a double-power law SFH on both sets of photometry simultaneously (i.e.~the combined CANDELS and CEERS photometric dataset), we find the best-fit model to be quenched (${\rm SFR} \sim 8~\msun~{\rm yr}^{-1}$), further indicating the importance of the mid-IR photometry in supporting a quenched SFH for UMG-28740.

One argument against a quenched SFH is the detection of Ly$\alpha$ emission. However, this may be explained by either AGN contribution or remnant star formation consistent with a post-starburst SED. Indeed, a recent study of a post-starburst galaxy at $z \sim 3$ has detected significant amounts of neutral gas from AGN driven outflows via ALMA \citep{Scholtz_2024}. Outflowing neutral gas around UMG-28740 could also contribute to the asymmetric Ly$\alpha$ profile measured in \S\ref{subsec:spectra}.

To explore the consistency of the observed Ly$\alpha$ profile with a quiescent model, we compare the observed and expected Ly$\alpha$ equivalent widths (EW) from different SED fits. Assuming no AGN contamination, we again use the `predict' function in \texttt{BAGPIPES} to estimate the EW of Ly$\alpha$ in the various SED fits. For the nearly quenched scenario, the SFR from all models without an AGN component ranges from 9.5~$\msun~{\rm yr}^{-1}$ to 32.2~$\msun~{\rm yr}^{-1}$. Measuring the EW of Ly$\alpha$ from the model SEDs (using the $\pm 1 \sigma$ spread in the quenched models) yields an EW$_\mathrm{Ly\alpha} = 16.5-35.1$\AA. We then estimate the EW of the observed Ly$\alpha$ line from Keck/DEIMOS. We find the continuum level just redward of the line profile and calculate the EW using a custom python code. Because the continuum level is so low, we conservatively call this measurement an upper limit on the true EW. Performing this analysis yields an observed EW$_\mathrm{Ly\alpha} > 34.0$\AA, within the range of the predicted values for the post-starburst SED. Hence, we cannot rule out the nearly quenched scenario by detection of Ly$\alpha$ emission alone. 

We perform the same analysis on the SEDs with AGN contribution. With AGN contribution we find the $\pm 1 \sigma$ spread for the quenched SEDs yields EW$_\mathrm{Ly\alpha} = 29.9-31.7$\AA, which is below the estimated observed Ly$\alpha$ EW. This would suggest that AGN contribution alone may not explain the presence of Ly$\alpha$ emission seen in the Keck/DEIMOS spectroscopy, and that likely there is some remnant star formation (even with the presence of AGN contamination). 

The best-fit SEDs for UMG-28740 yield two different SFH scenarios, a highly star-forming galaxy or a nearly quiescent one. However, due to the flux required by the {\it JWST}/NIRCam photometry in the mid-IR (F277W, F356W, F410M, and F444W) and the consistency in the EW of the observed and predicted Ly$\alpha$ emission line, we argue that the quenched SED is likely the true SFH of this object. 

\subsection{Comparison to Simulations}

\begin{figure}
    \centering
    \includegraphics[width=\linewidth]{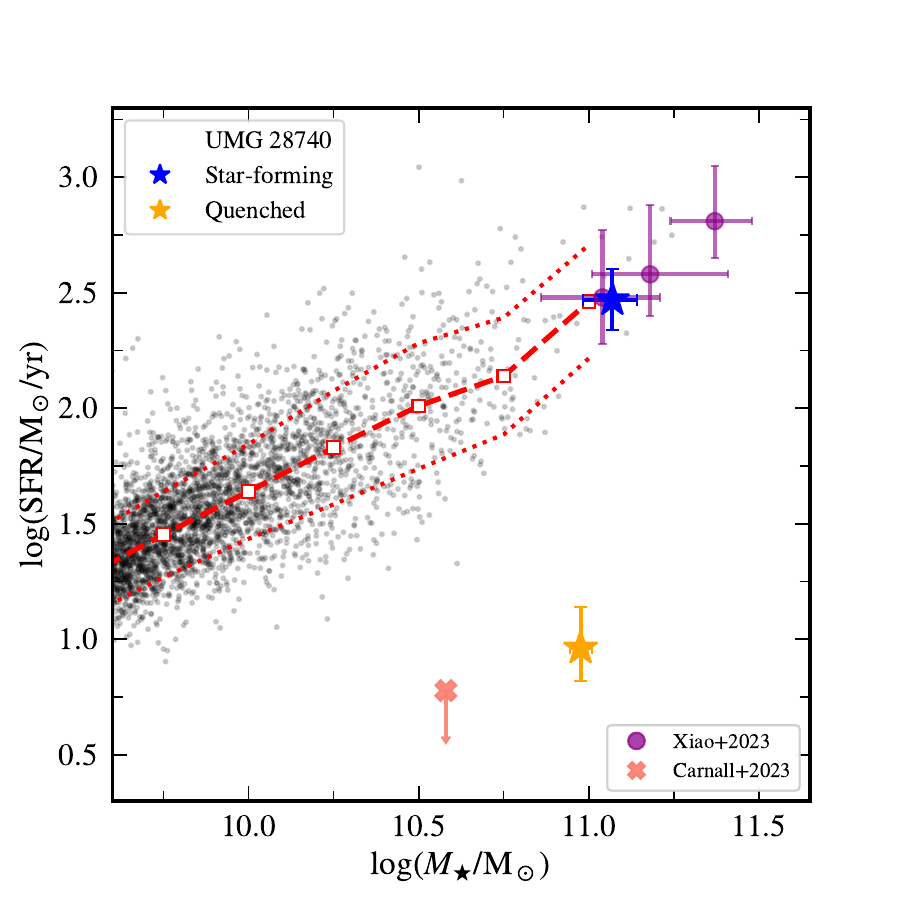}
		\caption{SFR versus stellar mass of UMG-28740 for the two scenarios, star-forming (blue star) and quenched (orange star). The black circles show galaxies from the Illustris TNG-300 simulation at $z = 5$. The red dashed line and white squares trace the median of the galaxies in the TNG sample, while the red dotted lines show the $\pm 1 \sigma$ spread. We also include other spectroscopically-confirmed massive galaxies at this epoch from the literature as purple circles \citep{Xiao_23} and a peach cross \citep{Carnall_2023_N}. 
  \label{fig:SFRvM}}
\end{figure}

The number density of UMGs at $z \sim 5$ is extremely low, especially for more quiescent systems \citep{Stefanon_2015, Davidzon_2017, Marsan_2022, Weaver_2023}. In addition, due to various observational challenges associated with observing these systems, there are only a handful of spectroscopically-confirmed UMGs at this epoch in the literature \citep[e.g.][]{AntwiDanso_2023,Xiao_23,Carnall_2023_N}. To compare the SFR of UMG-28740 to a large, unbiased sample of galaxies, we plot SFR versus stellar mass for UMG-28740 with simulated galaxies. Many simulations struggle to produce massive systems at high redshift, and most fail to quench these objects at early epochs. However, the Illustris TNG-300 simulation \citep{Nelson_2019} has been able to form massive systems and subsequently suppress SFR to reproduce quiescent UMGs up to $z \sim 3.5$ \citep{Forrest_2020a,Schreiber_2018,Glazebrook_2017}. Hence, we compare UMG-28740 to simulated galaxies in TNG-300 at $z = 5$ (see Fig.~\ref{fig:SFRvM}).

While TNG-300 reproduces massive quiescent systems at $z \sim 3.5$, as well as massive star-forming galaxies at earlier epochs \citep{Forrest_2020a}, TNG-300 fails to create massive and quiescent galaxies like UMG-28740 (assuming a quenched scenario) and one other spectroscopically-confirmed quenched system in the literature to date \citep{Carnall_2023_N} at $z \sim 5$. The measured SFR of UMG-28740 for the quenched scenario is $\sim 2$ dex below that of any other simulated galaxy at $\log(\mstar/\msun) \sim 11$. On the other hand, the TNG-300 simulation does reproduce massive, star-forming galaxies at this epoch, including those from \citet{Xiao_23} and the star-forming scenario of UMG-28740. The lack of massive quenched objects is likely a result of one of two possibilities: the TNG-300 simulation fails to catch the cessation of star formation at this epoch or that the size of TNG-300 is too small ($V = 302.6^{3}$ Mpc$^3$) to capture a system as rare as UMG-28740 and its protocluster environment. 

\begin{figure*}    
\centering
\includegraphics[width=\linewidth]{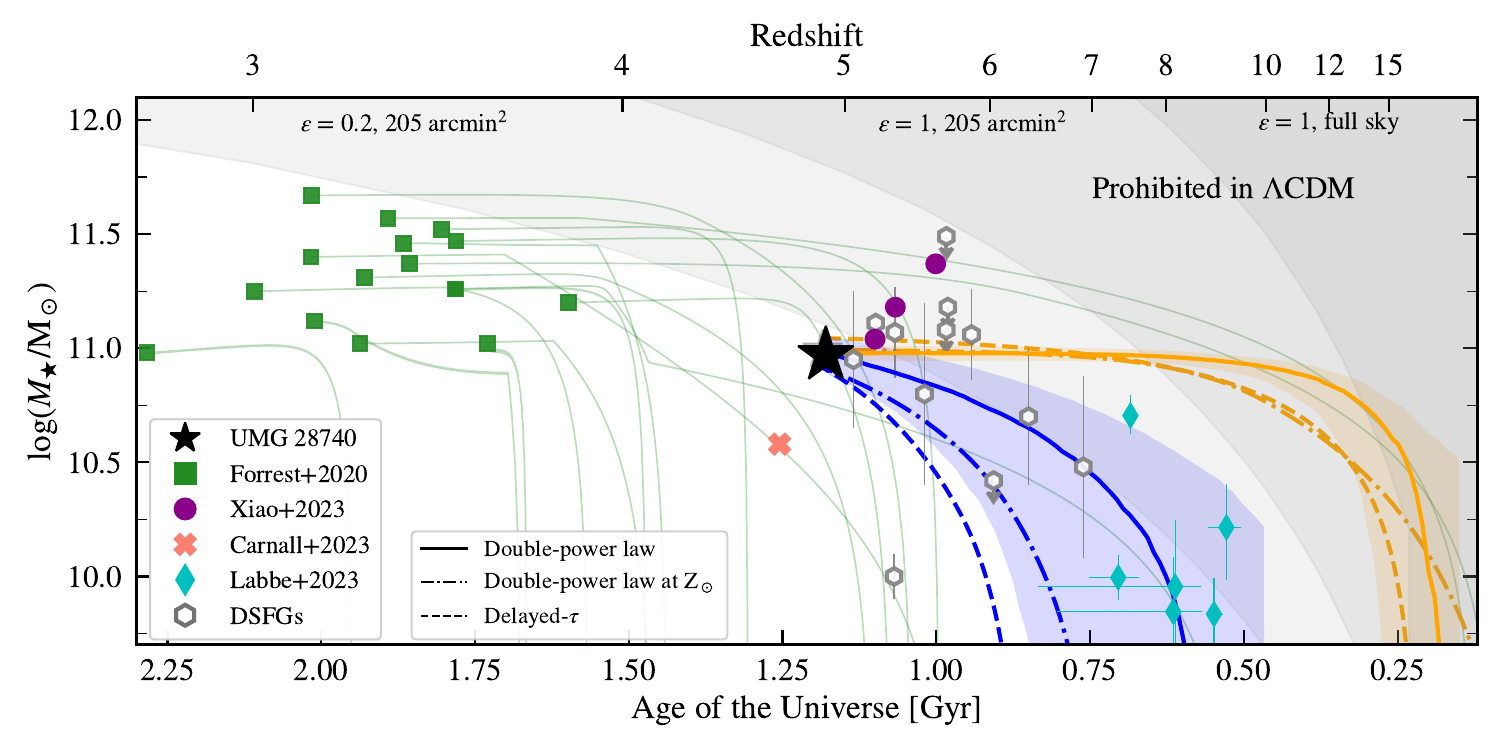}
		\caption{Mass assembly history of UMG-28740 for the best-fit star-forming models (blue lines) and quiescent models (orange lines). We include the double-power law fits (solid line), the fixed $Z={\rm Z}_{\odot}$ double-power law fits (dashed  dotted lines), and the delayed-$\tau$ fits (dotted lines). The $\pm 1\sigma$ uncertainties for the double-power law fits are shown as the shaded blue and orange regions. The green tracks and squares show the assembly histories for spectroscopically-confirmed UMGs at $z \sim 3$ from \citet{Forrest_2020b}. Spectroscopically-confirmed massive galaxies at $z \sim 5$ from \citet{Xiao_23} and \citep{Carnall_2023_N} are shown as purple circles and a peach cross, respectively. Finally, we include a sample of $z > 5$ DSFGs (grey open hexagons) from the literature and a sample of photometric massive galaxy candidates from \citet{Labbe_2023} (cyan diamonds). The grey shaded regions show the stellar mass prohibited by $\Lambda$CDM cosmology, following $\mstar^\mathrm{max} = \epsilon f_b {M_\mathrm{halo}}^\mathrm{max}$ where $\epsilon$ is the star formation efficiency and $f_b$ is the cosmic baryon fraction. We include three different thresholds, by assuming the area of the EGS at a 20$\%$ efficiency and a 100$\%$ efficiency as well as the area of the full sky and 100$\%$ efficiency. 
\label{fig:mass_evo}}
\end{figure*}

Recently, \citet{Xie_2024} used the GAEA semi-analytic model \citep{DeLucia_2024}, applied to a larger box size than that of TNG-300 ($V = 685^{3}$~Mpc$^3$), to study the formation of massive quiescent galaxies at early cosmic times. In summary, \citet{Xie_2024} find a larger relative number of quenched massive galaxies at $z \sim 4.5$, with a quenched fraction for systems with $\mstar = (0.8-1.2)\times 10^{11}~\msun$ of $f_{q} \sim 0.06$. \citet{Xie_2024} suggest that the primary quenching mechanism for massive galaxies at high redshift is accretion disk feedback via quasar winds. Yet even within the GAEA simulation UMG-28740 is exceptionally rare; galaxies as quiescent as the best-fit quenched model, with a specific SFR of $\sim 10^{-10}~{\rm yr}^{-1}$, have a number density of only $\sim 2 \times 10^{-7}$ per ${\rm cMpc}^{3}$ at $z \sim 5$ in GAEA \citep{DeLucia_2024}.

Observational studies suggest that massive quiescent galaxies at $z \sim 5$ exist at greater numbers than those predicted in these models (e.g. \citealt{Carnall_2023_Q}). However, spectroscopic detection of these systems has been \textit{nearly} unachievable prior to the launch of {\it JWST}. With {\it JWST} in operation, more quiescent massive systems are being uncovered (see \citealt{Carnall_2024} for 4 such systems at $3 < z <5$). These discoveries suggest systems such as UMG-28740 are not as unique as current simulations project, placing valuable constraints on modern models of galaxy formation.

\subsection{Mass Assembly History of UMG-28740}\label{subsec:LCDM}

To investigate the potential evolution of UMG-28740, in Figure~\ref{fig:mass_evo} we plot the mass assembly history corresponding to the best-fit star-forming and quiescent models from \texttt{BAGPIPES} (see Fig.~\ref{fig:competing_SFH}). We compare the mass growth of UMG-28740 to that of spectroscopically-confirmed UMGs at $z \sim 3.5$ from the Massive Ancient Galaxies At $z>3$ NEar-infrared (MAGAZ3NE) survey \citep{Forrest_2020a, Forrest_2020b, Forrest_2022}, a Keck/MOSFIRE effort that has successfully spectroscopically confirmed a large sample of star-forming and quenched UMGs at $z \sim 3.5$. We also include a sample of massive dusty star-forming galaxies (DSFGs) at $z > 5$ from the literature \citep{Riechers_2020,Williams_2019,Jin_2019,Zavala_2018,Pavesi_2018,Riechers_2017,Strandet_2017,Ma_2015,Cooray_2014,Riechers_2013,Walter_2012}, including the dusty star-forming UMGs spectroscopically confirmed by \citet{Xiao_23}. Finally, we also include the massive, spectroscopically-confirmed quiescent galaxy from \citet{Carnall_2023_N} and the photometric massive galaxy candidates from \citet{Labbe_2023}. Below we discuss the possible connections between UMG-28740 and these various populations.

For the mass assembly history predicted by the best-fit {star-forming} models, UMG-28740 is consistent with being the progenitor of the existing UMG population at $z \sim 3.5$, such that the ongoing star formation in the system will continue to grow its stellar mass at $3 \lesssim z \lesssim 5$, likely becoming similar to one of the more massive UMGs detected by the MAGAZ3NE survey. Similarly, the DSFG population at $z \sim 5$, including the galaxies identified by \citet{Xiao_23}, are also likely to evolve into members of the existing UMG population at $z \sim 3.5$ \citep{Forrest_2020b}. Looking towards higher $z$ for the best-fit star-forming models, the assembly history of UMG-28740 is consistent with the massive galaxy candidates from \citet{Labbe_2023} as potential progenitors, with UMG-28740 forming much of its stellar mass at $z \lesssim 8$.

On the other hand, assuming the more-likely quenched SFH, UMG-28740 will evolve into a lower-mass UMG at $z \sim 3.5$ ($\mstar \sim 10^{11}~\msun$). Interestingly, however, in this scenario UMG-28740 would not be included in the sample of known UMGs at $z \sim 3.5$. Due to the difficulty in detecting absorption lines in faint systems from the ground, MAGAZ3NE observations were limited to an AB magnitude of $K_{s} < 21.7$. In contrast, if we passively evolve the quenched model SED for UMG-28740 at $z = 4.8947$ to $z = 3.5$, we find that UMG-28740 would have an AB magnitude of $K_{s} = 22.7$, fainter than the selection limit for the MAGAZ3NE survey and indicating that existing samples of UMGs at intermediate redshift may be biased against the oldest systems (see \citealt{Forrest_2020b}). Instead, UMG-28740 would more likely resemble much fainter UMGs at $z \sim 3$, such as those that have been targeted with {\it JWST}/NIRSpec \citep{Nanyakkara_2024, Glazebrook_2024}. 

Towards high-$z$, the evolutionary history of UMG-28740 drastically changes assuming a quenched model. At $z \sim 5$, UMG-28740 is already an extreme system, more massive than the other spectroscopically-confirmed quiescent system at this epoch \citep{Carnall_2023_N}. Moreover, in contrast to the massive quiescent galaxy from \citet{Carnall_2023_N}, which formed its mass at $z \sim 7$, we find UMG-28740 to already be quenching by $z \sim 8$, with the bulk of its stellar mass formed at much earlier cosmic time ($z > 10$). Based on the quenched SED fits, the mass assembly of UMG-28740 is inconsistent with the DSFG population at $z > 5$ and the massive galaxy candidates from \citet{Labbe_2023}, instead forming a more extreme amount of mass at very high redshift. 

The formation of such massive galaxies at high redshift is a highly active area of study thanks to a wealth of new observations from {\it JWST}. For example, early {\textit{JWST}} imaging campaigns discovered a surplus of massive galaxy candidates at $z = 8-14$ that stand in tension with predictions from $\Lambda$CDM \citep{Casey_2023, Labbe_2023, BK_2023}. These results have sparked a myriad of questions and scrutiny in our current understanding of the physical processes driving galaxy evolution, including those that underpin SED fitting codes. The extreme build-up of mass required to assemble massive galaxies suggests intense bursts of star formation that convert $>20\%$ of the available baryons into stars \citep{Steinhardt_2016, BK_2023, Sun_2023, Casey_2023, Xiao_23, AntwiDanso_2023}.

To investigate the star formation efficiency of UMG-28740, we plot the limiting mass as imposed by a $\Lambda$CDM cosmology (shown as the grey shaded regions in Fig.~\ref{fig:mass_evo}). To do this, we follow the methodology from \citet{BK_2023}, first calculating the number density of halos of a given mass as a function of redshift using \texttt{hmf}\footnote{\href{https://hmf.readthedocs.io}{https://hmf.readthedocs.io}}, a python package that calculates the halo mass function at different redshifts \citep{Murray_2014}. We then find the maximum halo mass, given a survey volume, for which we would observe $1$ such system. To convert to stellar mass, we follow $\mstar^\mathrm{max} = \epsilon f_b  {M_\mathrm{halo}}^\mathrm{max}$, where $f_b$ is the cosmic baryon fraction and $\epsilon$ is the efficiency of converting baryons into stars. We perform this calculation for the area of the CANDELS footprint within the EGS (205 arcmin$^2$, \citealt{Stefanon_2017}) using a line-of-sight depth of $\Delta z = 2$ and an efficiency of 20$\%$ ($\epsilon = 0.2$) and 100$\%$ ($\epsilon = 1.0$). We also calculate the maximum stellar mass for observations across the full sky with an efficiency of 100$\%$, which is the upper limit for what is allowed in a $\Lambda$CDM cosmology. 

At $z \sim 5$, UMG-28740 has a stellar mass consistent with a relatively high star formation efficiency ($\sim 20\%$), comparable to that found in larger samples of massive galaxies at high redshift \citep[e.g.][]{Chworowsky_2023}. In contrast, recent analyses of massive star-forming galaxies at $z > 5$ find even more extreme star formation efficiencies \citep[$\epsilon \sim 0.5$,][]{Xiao_23, Casey_2023}. If UMG-28740 is quiescent at $z \sim 5$, then its evolution is also consistent with an elevated formation efficiency at higher $z$. As shown in Figure~\ref{fig:mass_evo}, at $z \gtrsim 8$ where the quenched model for UMG-28740 is still star forming, the implied formation efficiency is on the order of (and potentially in excess of) unity, consistent with the expectations from models of feedback-free star formation \citep{Dekel_2023}. Moreover, for the quenched model, UMG-28740 reaches significant tension with $\Lambda$CDM at even earlier times, exceeding $\epsilon = 1$ across the area of the full sky at $z \sim 15$. We tested this tension by fitting the CEERS photometry with a model set to high metallicity ($Z = 2 Z_\odot$), to further explore the potential degeneracy between stellar age and metallicity. 
Even at super-solar abundances, however, we find that the formation time is only reduced by $0.15$~Gyr, still in conflict with $\Lambda$CDM at high redshifts. This suggests UMG-28740 formed its stars at nearly impossible rates as currently predicted by theoretical models.

\subsection{RUBIES-EGS-QG-1}\label{subsec:Rubies}
While we finalized this manuscript, \citet{deGraaff_2024} presented a {\it JWST}/NIRSpec PRISM and G395M spectrum of UMG-28740 (identified as RUBIES-EGS-QG-1 in \citealt{deGraaff_2024}). The spectroscopic observations were collected in March 2024, with coverage of the Balmer break and the H$\alpha$ emission line. This effort was part of the RUBIES survey (GO-4233; PIs A. de Graaff and G. Brammer), which uses {\it JWST}/NIRSpec to observe galaxies in the EGS and UDS fields. In this section, we discuss the results of this independent work and compare those to our work.

The physical properties of RUBIES-EGS-QG-1 were measured from SED fitting of the {\it JWST}/NIRSpec PRISM spectrum, as well as photometry from {\it HST}/WFC3 (F125W, F140W, and F160W), {\it HST}/ACS (F814W and F606W), and {\it JWST}/NIRCam (F115W, F150W,
F200W, F277W, F356W, F444W, and F410M). The spectrum  and photometry were fit simultaneously via \texttt{Prospector} \citep{Leja_2017}, assuming a Chabrier IMF \citep{Chabrier_2003}, stellar population models from Flexible Stellar Population Synthesis \citep[FSPS,][]{Conroy_2010,Conroy_2009},  the MILES spectral library \citep{Sanchez_2006}, and MESA Isochrones and stellar tracks \citep[MIST,][]{Choi_2016,Dotter_2016}. With \texttt{Prospector}, \citet{deGraaff_2024} explored four different star formation history models -- a delayed-$\tau$, rising, and fixed $Z = Z_\odot$ model as well as a SFH comprised of 14 bins of constant star formation. The resulting best-fit stellar mass is in agreement with our results, yielding $\mstar = 8.7^{+0.5}_{-0.4} \times 10^{10}~\msun$. With inclusion of mid-IR spectroscopy from {\it JWST}/NIRSpec, \citet{deGraaff_2024} pin down the SFR in the last 100 Myr as ${\rm SFR}_{100} = 1.9^{+1.5}_{-1.0}~\msun~{\rm yr}^{-1}$, as the first work to spectroscopically confirm its quiescence. This independent work uses a separate SED modeling code and includes a high-quality spectrum that probes key spectral features, yet the final results are in excellent agreement with our quenched model. 

A remarkable note between the results presented in \citet{deGraaff_2024} and this manuscript is the detection of Ly$\alpha$ emission in a high-redshift quiescent galaxy. The observed spectrum and SED fit from \citet{deGraaff_2024} did not include a Ly$\alpha$ emission line, potentially due to poor resolution at the blue end of the {\it JWST}/NIRSpec PRISM spectrum and the low SFR in the SED models. We investigate the potential origins of this Ly$\alpha$ emission by estimating the EW of Ly$\alpha$ for a best-fit model SED to the observed photometry given SFR $= 6 ~\rm \msun yr^{-1}$ (the highest SFR in the last 10 Myr for any model in \citealt{deGraaff_2024}). We find the resulting Ly$\alpha$ EW to be $18.4^{+4.6}_{-2.5}$ \AA, well below the limiting EW given by our observations (EW$_\mathrm{Ly\alpha} > 34.0$ \AA, see \S\ref{sec:SFH}). This suggests that star formation from this galaxy alone may not be the cause of this Ly$\alpha$ emission, but rather there could be additional contribution from an AGN or the surrounding circumgalactic medium. In addition, \citet{deGraaff_2024} measured a slightly lower redshift via emission line modelling of [O{\scriptsize III}], [N{\scriptsize II}], [S{\scriptsize II}], and H$\alpha$ ($z_\mathrm{spec} = 4.8963$) such that the peak Ly$\alpha$ emission is redshifted by 479.67 km s$^{-1}$. Taken together with the asymmetric profile discussed in \S\ref{subsec:spectra}, this could be a sign of a significant amount of neutral gas surrounding the galaxy or AGN-driven galactic outflows.

The complementary analysis from \citet{deGraaff_2024}, taken together with the conclusions of our SED fitting, highlights the importance of high-quality mid-IR photometry -- currently only accessible via {\it JWST}/NIRCam -- in understanding systems like UMG-28740. Only with such data can we accurately measure key physical properties of $z \gtrsim 4$ galaxies. This example (UMG-28740) demonstrates the importance of {\it JWST} as we continue to search and study the rarest and most massive systems in the Universe.

\section{Conclusions} \label{sec:conc}

In this work, we present spectroscopic confirmation of an ultra-massive galaxy (UMG, ID 28740) with $\log(\mstar/\msun) = 10.98\pm0.07$ at $z_\mathrm{spec} = 4.8947$ based on deep spectroscopy with Keck/DEIMOS. UMG-28740 is one of the most massive spectroscopically-confirmed galaxies at $z \sim 5$ and lies within one of the most extreme overdensities in the EGS. Here, we summarize the main results of our analysis:
\begin{itemize}

    \item We observe an asymmetric Ly$\alpha$ emission line profile with a relatively high asymmetry parameter of $A_f = 3.56$, suggesting the presence of neutral gas within or surrounding the galaxy. This neutral component could be due to gas in the interstellar medium, circumgalactic medium, or potentially associated with AGN-driven outflows. 

    \item Using the spectroscopic redshift, we run a large suite of SED modeling (assuming various SFHs and using two SED fitting codes, \texttt{BAGPIPES} and \texttt{FAST++}) on two sets of photometric data from CANDELS \citep{Stefanon_2017} and CEERS \citep{Bagley_2023}. Regardless of the SED code, set of free parameters, or SFH assumed, the stellar mass of UMG-28740 is tightly constrained, with a median mass of $\log(\mstar/\msun) = 10.98$ and a standard deviation of $\pm0.07$ between all runs.  

    \item We find a discrepancy between the measured SFR and photometric dataset assumed, with multiple SED fits to the CANDELS photometry yielding a highly star-forming object. However, when fitting to the {\it JWST}/NIRCam photometry from CEERS, UMG-28740 is nearly quiescent, with SFR $<32.2~\msun~{\rm yr}^{-1}$. Futhermore, while the quenched SED is consistent with both sets of photometry, the predicted flux from the best-fit star-forming SED under-predicts the measured flux at F277W and over-predicts the measured flux at F356W, F410M, and F444W; suggesting a quenched scenario is more likely. Finally, we show the presence of Ly$\alpha$ does not rule out a post-starburst, nearly quenched scenario.

    \item We find UMG-28740 to be the most massive member of a protocluster environment at $z \sim 5$, with the surface density of galaxies in the region around UMG-28740 roughly $10\times$ higher than the average coeval field surface density. Within $R_\mathrm{proj} = 3.44$ cMpc of UMG-28740, we find 26 photometric members with $\mstar > 10^{9}~\msun$ and 160 photometric members with $\mstar > 10^{8}~\msun$ at $4.5 < z < 5.5$, using the CANDELS and CEERS photometric catalogs, respectively. Finally, we find four additional massive galaxies ($\mstar > 10^{10.5}~\msun$) within $0.29-1.13$~cMpc of UMG-28740 in the CANDELS photometric catalog. 

    \item Comparing the SFR of UMG-28740 to the Illustris TNG-300 simulation box at $z=5$  shows that, if quenched, UMG-28740 is extremely rare -- with a SFR $\sim 2$ dex lower than any simulated galaxy at $\log(\mstar/\msun) \gtrsim 11$. 

    \item We find UMG-28740 has a high star formation efficiency ($\sim 20\%$) at $z \sim 5$. Assuming a quenched model, this efficiency exceeds $\epsilon = 1$ for the area of the full sky, suggesting UMG-28740 would need to form its stars at faster rates than predicted by theoretical models.  
\end{itemize}

\section*{Acknowledgements}

The authors wish to recognize and acknowledge the very significant cultural role and reverence that the summit of Maunakea has always had within the indigenous Hawaiian community. We are most fortunate to have the opportunity to conduct observations from this mountain. 
This work is based in part on observations taken by the CANDELS Multi-Cycle Treasury Program with the NASA/ESA {\it Hubble Space Telescope}, which is operated by the Association of Universities for Research in Astronomy, Inc., under NASA contract NAS5-26555.
This work is also based in part on observations taken by the Cosmic Evolution Early Release Science Survey (CEERS) team with the NASA/ESA/CSA {\it JWST} from the Space Telescope Science Institute, which is operated by the Association of Universities for Research in Astronomy, Inc., under NASA contract NAS 5-03127.
SMUS and MCC acknowledge support from the National Science Foundation through grant AST-1815475. 
BF acknowledges support from JWST-GO-02913.001-A. 
GW gratefully acknowledges support from the National Science Foundation through grant AST-2205189 and from HST program number GO-16300. 
This research was supported by the International Space Science Institute (ISSI) in Bern, through ISSI International Team project ``Understanding the evolution and transitioning of distant proto-clusters into clusters".
We also gratefully acknowledge the Lorentz Center in Leiden (NL) for facilitating discussions related to this project.
MCC thanks Michael Boylan-Kolchin, Kristin Turney, and Lucas Cooper for helpful discussions related to this work. 
SMUS thanks Adam Carnall for additional support with \texttt{BAGPIPES} functionality. 
Finally, we thank the anonymous referee for their insightful comments and suggestions that improved this work. 

\bibliographystyle{mnras}
\bibliography{biblio}


\label{lastpage}
\end{document}